\documentclass[twocolumn,prl]{revtex4-1}
\usepackage{graphicx}
\usepackage{epstopdf}
\usepackage{dcolumn}
\usepackage{amsmath}
\usepackage{multirow} 
\usepackage[active]{srcltx}
\usepackage{hyperref}

\begin{document}

\newcommand{\K}{{\vec k}}

\title{Substitutional nickel impurities in diamond: decoherence-free subspaces for quantum information processing}

\author{T. Chanier, C. E. Pryor, and M. E. Flatt\'e}
\affiliation{Optical Science and Technology Center and Department of Physics and Astronomy, University of Iowa, Iowa City, Iowa 52242, USA}

\begin{abstract}
The electronic and magnetic properties of a neutral substitutional nickel (Ni$_s^0$) impurity in diamond are studied using density functional theory in the generalized gradient approximation.  The  spin-one ground state consists of two  electrons with parallel spins, one located on the nickel ion in the $3d^9$  configuration and the other distributed among the nearest-neighbor carbons. The exchange interaction between these spins is due to $p-d$ hybridization and is controllable with compressive hydrostatic or uniaxial strain, and for sufficient strain the antiparallel spin configuration becomes the ground state. 
Hence, the Ni impurity forms a controllable two-electron exchange-coupled system that should be a robust qubit for solid-state quantum information processing.
\end{abstract}

\maketitle

Recent advances in single-spin control of electron spins in quantum dots\cite{Hanson2007} and spins associated with single dopants in semiconductors\cite{Koenraad2011} suggest such systems could make good qubits for quantum information processing\cite{Awschalom2002}.
Spin centers based on single impurities or impurity complexes in diamond have been extensively explored as qubit candidates due to effective optical access and extremely long room-temperature spin coherence times\cite{Jelezko2006,Fuchs2008,Hanson2008,Neumann2009,Balasubramanian2009}. 
Although a multitude of spin centers have been observed or predicted to occur in diamond, including transition metal dopants like nickel, 
the spin-one nitrogen-vacancy (NV$^-$) center has drawn the most attention due to the successful demonstration of optical spin initialization and readout\cite{Jelezko2006,Childress2006,Epstein2005,Hanson2008b,Jiang2009} and spin manipulation via interaction with a neighboring nitrogen nuclear spin\cite{Jelezko2004b,Neumann2008} or electron spin\cite{Hanson2006}. 
Ni can be introduced in diamond by ion implantation or during chemical vapor deposition \cite{Wolfer2010,Orwa2010}. 
Characterization by optical spectroscopy, electron paramagnetic resonance (EPR) and magnetic circular dichroism (MCD) \cite{Isoya1990,Iakoubovskii2004,Isoya1990b,Mason1999} has revealed the existence of several Ni related optical centers, in particular a 1.4 eV optical feature which has been attributed to a Ni$^+$ interstitial with a  $3d^9$ configuration and spin $S=\frac{1}{2}$. To our knowledge, no experimental study of the  substitutional Ni spin center has been  reported.  

In other material systems the coherence times and fidelity for quantum operations of qubits has been improved dramatically through the use of decoherence-free subspaces of exchange-coupled spins to form a qubit; one example is the use of the spin singlet state and the $S_z=0$ state of a spin triplet to form a decoherence-free subspace for two electrons confined to two neighboring quantum dots\cite{Levy2002,Petta2005}.
These approaches rely on control of the exchange interaction between two spins, such as by using an electrical gate to modulate the electronic hopping from one quantum dot to another. Electric control of the exchange interaction is challenging for a spin center in diamond, for the typical size of the electronic wave functions is very small, and gating technology is not well advanced. Strain provides an alternate mechanism for controlling a spin center, as was recently shown via electrically-detected magnetic resonance induced by strain control of the hyperfine constant of a $^{31}$P$^+$ donor in strained Si\cite{Dreher2011}.

Here we show, using density functional theory calculations, that the substitutional nickel impurity in diamond can be understood as an exchange-coupled system of two electron spins: one localized on the nickel ion and one delocalized on the four nearest-neighbor (NN) carbon atoms. The electronic configuration of Ni is unambiguously determined by $p$--$d$ hybridization  between the Ni $3d$ and NN carbon $2p$ levels.  Although the ground state at ambient pressure for the neutral nickel impurity has been predicted to be a spin-one center \cite{Larico2009}, the spin-zero state is nearly degenerate, and can be made degenerate through the application of reasonable compressive hydrostatic or uniaxial strain. To reduce to an effective two-state decoherence free subspace, similar to that implemented for quantum dots \cite{Petta2005}, the $S_z=\pm 1$ triplet states ($T_{\pm}$) could be split off by a magnetic field. For the nickel ion, strain modulation could be used to manipulate the energy splitting between the singlet ($S$) and the remaining $S_z=0$ triplet state ($T_0$),  instead of the electrostatic gating used for double quantum dots \cite{Petta2005}. As the environments differ for the two electron spins that combine to form the $S$ or $T_0$ states, the slightly different $g$ factors we find for these two spins provide an orthogonal axis of control in the effective two-state $S$/$T_0$ subspace of the nickel ion.

The calculations were performed with the scalar relativistic version of the full potential local orbital FPLO9.00-33 code \cite{Koepernik1999} using the (spin-polarized) generalized gradient approximation ((S)GGA) with the parametrization of Perdew, Burke
and Ernzerhof \cite{Perdew1996}. 
The convergence of the results with respect to ${\bf k}$-space integrals was carefully checked, and 8x8x8 = 512 {\bf k}-points were sufficient. 
We used a 64-atom NiC$_{63}$ supercell  corresponding to a $2\times 2 \times 2$ multiple of the (non-fundamental) cubic unit cell of diamond with Ni substituted for one carbon. 
The supercell size was fixed to correspond to the experimental lattice constant of diamond $a_0=3.5668$\AA \cite{Ripley1944}, and all the 
atomic positions within the supercell were allowed to relax with a precision of 1 meV/\AA.   
We considered two possible symmetries for the Ni impurity,  $T_d$ with four identical NN carbons, and $C_{3v}$ in which a trigonal distortion along the [111] axis is allowed.
We found that the relaxation primarily involves the NN carbons which moved by 14.2\% while  further neighbors moved by  $<$ 1.1\%. 
The $C_{3v}$ relaxation along [111] was less than 0.05\% and so we considered only $T_d$ in subsequent calculations. 
The electron occupation number of Ni was 26.9 and the  total magnetization of the ground state was 2.0 $\mu_\mathrm{B}$, with approximately 0.8 $\mu_\mathrm{B}$  localized on the Ni, and the rest distributed among the NN carbons.

\begin{figure}[h!]
\begin{center}
\includegraphics[width = 90mm]{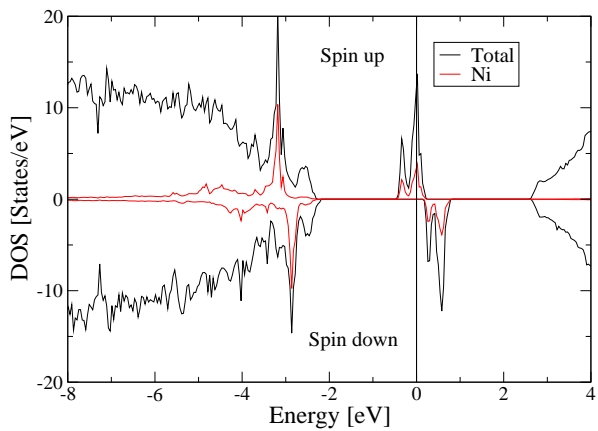}\\[-0.4cm]
\caption{Total and Ni partial SGGA DOS of a NiC$_{63}$ supercell.}\label{DOSnoStr}
\end{center}
\end{figure}

Fig. \ref{DOSnoStr} shows the total and Ni partial SGGA density of states (DOS) of the NiC$_{63}$ supercell. 
The calculated direct band gap of 4.9 eV and  indirect  gap of 4.8 eV are both smaller than the experimental  gap $E_G^{Ind}=5.47$ eV, due to the gap underestimation of the GGA functional \cite{Filippi1994}. 
The Ni $3d$ levels are split by the crystal field into doubly degenerate $e_g$ states and a higher energy triplet of  $t_{2g}$ levels. 
The spin up and spin down $e_g$ levels are localized on the Ni site, giving rise to peaks approximately  1 eV below the valence band maximum (VBM) and  separated by a Hund exchange splitting $J^H=0.4$ eV. 
The $t_{2g}$ levels strongly hybridize with the valence band  (predominantly  $2p$ character), forming a partially filled and a totally empty bound state at  2.5 and 3.0 eV above the VBM respectively, corresponding to the spin up and spin down antibonding levels as described below. 
We note the presence of a ferromagnetic (FM) spin splitting of the valence band corresponding to the strong hybridization limit similar to the case of Mn$^{2+}$ in ZnO or Mn$^{3+}$ in GaN \cite{Dietl2008,Chanier2009}.

The separate electronic configurations of the Ni and the neighboring carbon atoms clarify the electronic nature of the impurity. 
Ni is in the positively charged [Ar]$4s^03d^9$ configuration Ni$^+$ with a spin $S_1=\frac{1}{2}$. Nickel's positive charge induces a single electron in the surrounding diamond that is distributed among the dangling bonds of the NN carbons.
This addition of the induced electron to the four dangling bonds, each with $2s2p$ character, results in a $2s^22p^3$ configuration with   $S_2=\frac{1}{2}$.

\begin{figure}[h!]
\begin{center}
\includegraphics[width = 77mm]{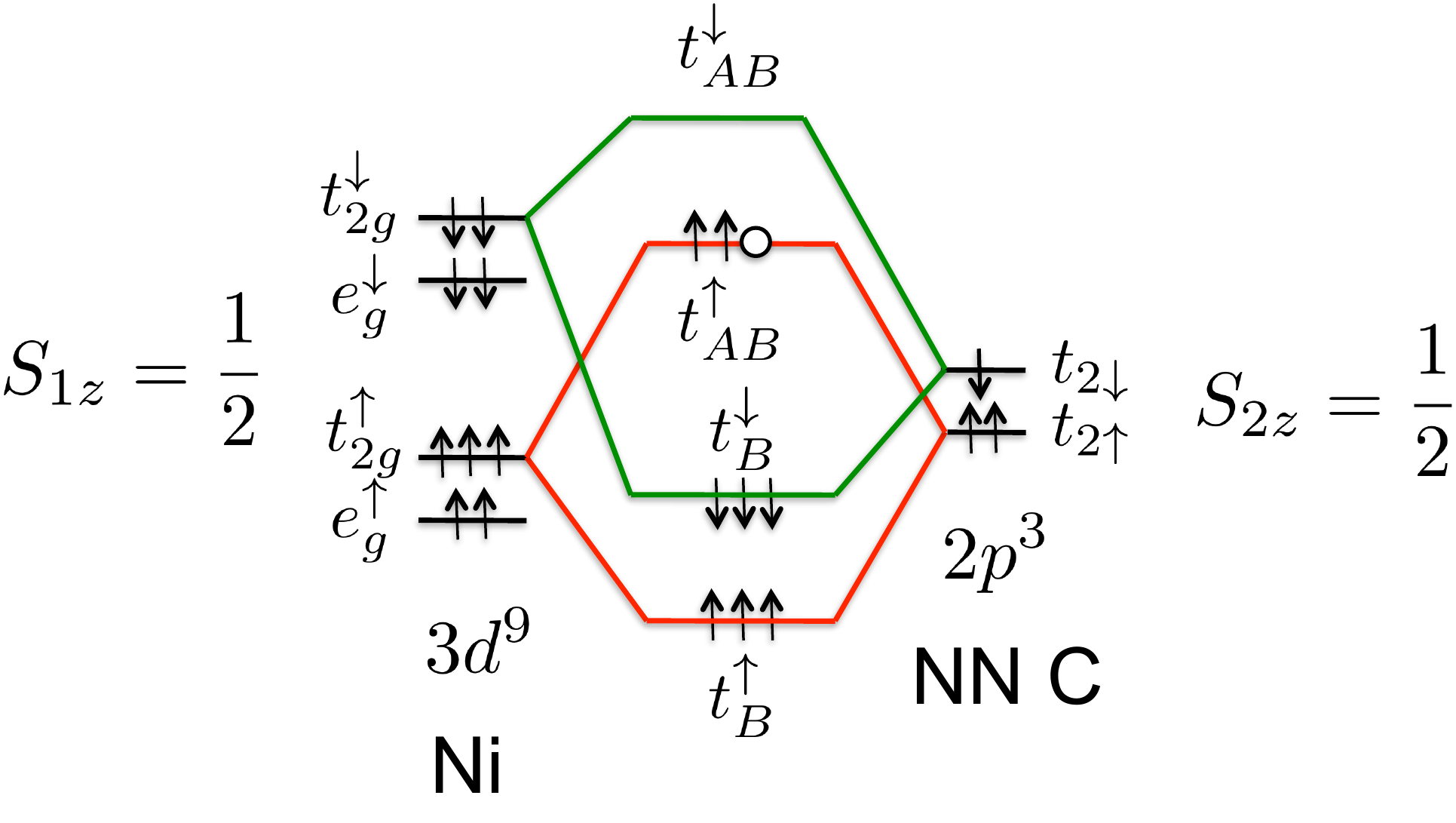}
\caption{Hybridization between the Ni $e_g$ and $t_{2g}$ $3d$ levels and the $2p$-derived nearest-neighbor carbon dangling bond with a ferromagnetic alignment  providing a total spin $S_T=1$ (triplet state $T$).\\[-1.4cm]}\label{pdFMcoupling}
\end{center}
\end{figure}

The levels of the Ni$^+$ and NN carbons hybridize, as shown in Fig.  \ref{pdFMcoupling}. 
The $2p$ derived $t_2$ defect level  hybridizes with the Ni $t_{2g}$ levels to form bonding ($t_B^{\uparrow/\downarrow}$) and antibonding  ($t_{AB}^{\uparrow/\downarrow}$) levels. 
This $p$--$d$ hybridization picture identifies the origin of the calculated SGGA DOS (Fig. \ref{DOSnoStr}) as due to a FM interaction between $S_1$ and $S_2$.   

\begin{figure}[h!]
\begin{center}\includegraphics[width = 85mm]{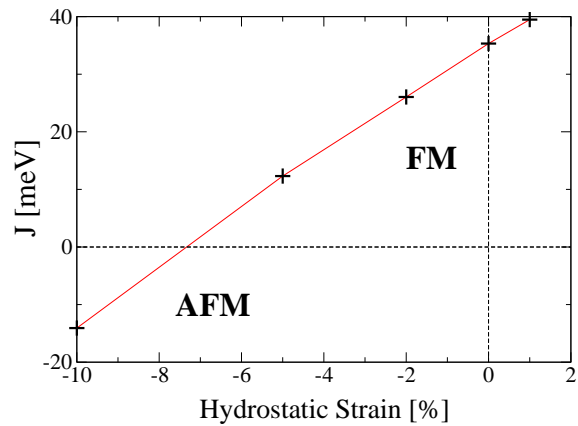}\\[-0.2cm]
\caption{Heisenberg exchange coupling between the two spins $S_1$ and $S_2$ calculated as a function of hydrostatic strain.}\label{JvsStr}
\end{center}
\end{figure}

Strain dramatically modifies the  exchange coupling between the  spins  of the Ni$^+$ and the surrounding carbons. The energy difference between the FM and AFM arrangements of the two spins  was equated to the total energy difference between the SGGA and GGA calculations $\Delta E=(E_{SGGA}-E_{GGA})/2=-({J}/{2}) S_T(S_T+1)$, with $S_T=S_1+S_2$ and the Hamiltonian $H=-2J \mathbf{S}_1 \cdot \mathbf{S}_2$. Hydrostatic strain was included by changing the size of the supercell, within which the atomic positions were allowed to relax.  For  [001] uniaxial strain, we fixed the atomic positions  to their calculated positions for  the unstrained system
and rescaled the supercell anisotropically according to the Poisson ratio \cite{Grimsditch1975}.

Fig. \ref{JvsStr} shows the exchange coupling as a function of hydrostatic strain, where J $>$ 0 and  J $<$ 0 correspond to  FM and AFM coupling respectively. The ground state has an AFM spin alignment  for $e_{H}<-7$ \%.
As a check, we used an initial configuration of 1 $\mu_\mathrm{B}$ on the Ni and -1 $\mu_\mathrm{B}$  on the NN carbons to locate the low-spin (non-ground-state) AFM solution with the SGGA functional for $e_H=0$. The total energy calculated for this solution is within 3~meV of the nonmagnetic GGA solution, which is within our estimated error. The SGGA AFM ground state for $e_H=-10$ \% is similarly degenerate with the GGA result, and the calculated total magnetization is nearly zero ($\sim 0.1\ \mu_\mathrm{B}$).
Under [001] uniaxial strain, the ground state becomes AFM for a compressive strain $e_{[001]}<-19$ \%. The existence of a pressure induced transition between a high-spin and low-spin state is also observed in transition-metal-ion compounds, where it was shown 
theoretically to be due to a competition between localization (induced by the Hund's exchange coupling which favors the high-spin state) and a tendency towards delocalization induced by the crystal field (which favors the low-spin state) \cite{Bari1972}.

\begin{figure}[h!]
\includegraphics[width =80mm]{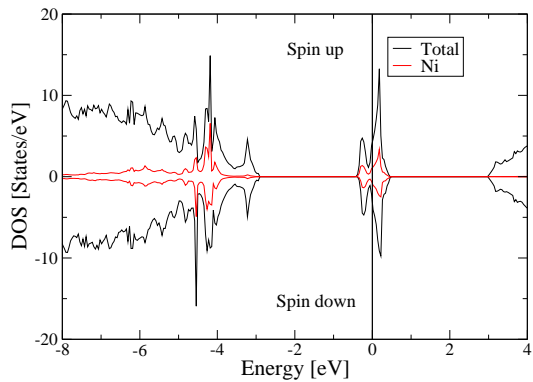}\\[-0.4cm]
\caption{Total and Ni partial SGGA DOS of a NiC$_{63}$ supercell with a compressive hydrostatic strain $e_{H}=-10$\%.}\label{DOSwithStr}
\end{figure}

\begin{figure}[h!]
\includegraphics[width = 80mm]{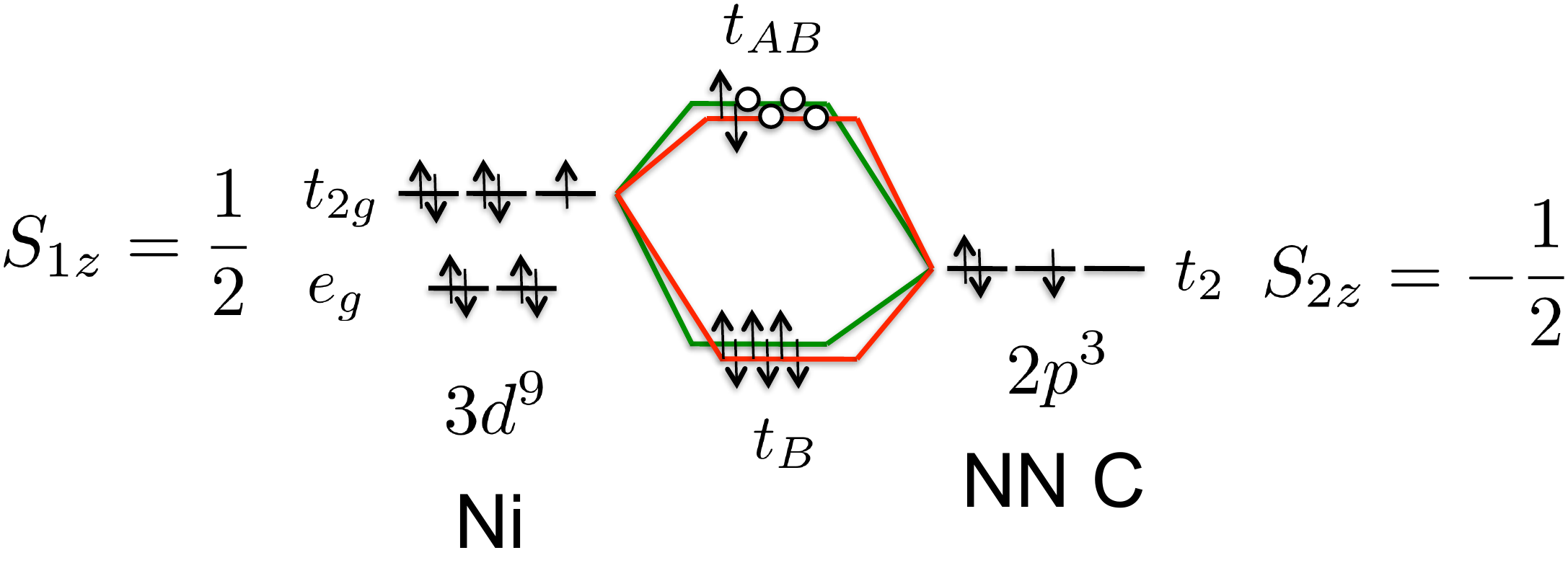}
\caption{Electronic configuration corresponding to an AFM alignment of $S_1$ and $S_2$, providing a total spin $S_T=0$ (singlet state $S$).}\label{pdAFMcoupling}
\end{figure}

Fig. \ref{DOSwithStr} presents the total and Ni partial SGGA DOS of the AFM  ground state with $e_{H}=-10$ \%.
We found that the band gap increases to  5.8 eV. The spin up and spin down $e_g$ levels are now spin degenerate ($J^H=0$), and there is a hybridization between the Ni $t_{2g}$ and the NN C $2p$ levels, leading to the formation of bonding and antibonding levels. The spin-degenerate antibonding levels are partially filled at the Fermi level and correspond to a bound state in the gap  3.4 eV above the VBM. 
Fig. \ref{pdAFMcoupling} gives a schematic representation of the $p$--$d$ hybridization model under strain. The crystal field splits the Ni $3d$ level into lower $e_g$ and upper $t_{2g}$ levels which are now spin degenerate ($J^H=0$). To explain the non-magnetic DOS presented in Fig. \ref{DOSwithStr}, we need to take into account an AFM interaction between the two spins $S_1$ and $S_2$.

The Ni$_s^0$ system can be described in the singlet $S$ (AFM, $S_T=0$) - triplet $T$ (FM, $S_T=1$) basis by  the Hamiltonian $H=-2J(\{e\})\mathbf{S}_1 \cdot \mathbf{S}_2+\mu_\mathrm{B}(g_1S_{1z}+g_2S_{2z})B_z$. With the $T_{\pm}$ states split off by an applied magnetic field $B_z$, we have a two-level $S/T_0$ system (qubit). In the $S/T_0$ subspace, $H$ can be written as follows:
\begin{equation}
H =\left(\begin{array}{cc} \mathcal{E}_S(\{e\})& \Delta \mathcal{G}^z\\ \Delta \mathcal{G}^z&0\end{array}\right),\label{2stateham}
\end{equation}
with $\mathcal{E}_S(\{e\})=2J(\{e\})$ the strain-controlled exchange splitting ($e.g.$ Fig. \ref{JvsStr}), $\Delta \mathcal{G}^z=(\mu_\mathrm{B} B_z\Delta g)/4$ and the basis states $|S\rangle=\frac{1}{\sqrt{2}}(|\uparrow_{\rm Ni}\downarrow_{\rm C}\rangle-|\downarrow_{\rm Ni}\uparrow_{\rm C}\rangle)$, $|T_0\rangle=\frac{1}{\sqrt{2}}(|\uparrow_{\rm Ni}\downarrow_{\rm C}\rangle$ $+|\downarrow_{\rm Ni}\uparrow_{\rm C}\rangle)$. Rapid dynamical spin exchange occurs between the Ni and C sites, with a speed that can be estimated as $\sim 1$~fs from the $\sim 3$~eV splitting between bonding and antibonding $t_{2g}$ states in Fig.~\ref{pdFMcoupling}.

The singlet-triplet mixing $\Delta \mathcal{G}^z$ depends on the difference in the Land\'e factors $\Delta g=g_1-g_2$ between the spin $S_1$ localized on the Ni and the spin $S_2$ located on the NN carbon atoms. Applying Hund's rule to the Ni spin we obtain a $g_1$ factor of 1.33, whereas the spin on the NN carbons is delocalized and thus should be well described by a free-electron $g_2$ factor of $2$. We obtain a finite difference $\Delta g \sim -0.66$. The Ni$_s^0$ qubit, a two exchange-coupled spin-$\frac{1}{2}$ system, can therefore be used to develop an universal quantum computer (QC) architecture \cite{Levy2002}. The spins $S_1$ and $S_2$ form nearly decoherence free subspaces (DFS) due to the long spin lifetime in diamond. Based on these DFSs,  a solid-state QC architecture based on Ni$^0_s$ qubit should be immune from collective decoherence mechanism.

Generation of sufficient strain to make the FM ($T_0$) and AFM ($S$) states degenerate  would require a stress of $\sim 80$~GPa, which is well below the $>300$~GPa maximum of a diamond anvil cell\cite{Hemley2008}; such strains and stresses are also typical in nanoscale pseudomorphic structures, such as InAs/InSb heterostructure nanowires \cite{Caroff2008}. Once the $S$ and $T_0$ states are degenerate, high-speed manipulation of the two-state system with modulated strain can be done with strain modulation amplitudes orders of magnitude less.  To achieve an energy splitting corresponding to a $\sim 1$~GHz precession frequency, a strain modulation of only $10^{-5}$ would be required, which is three orders of magnitude smaller than that used in Ref. \cite{Dreher2011}. High-$Q$ nanomechanical diamond resonators operating above $1$~GHz have already been demonstrated \cite{Gaidarzhy2007}.

\begin{figure}[h!]
\begin{tabular}{|c|c|}
\hline & \\[-0.2cm]
\includegraphics[width = 32mm]{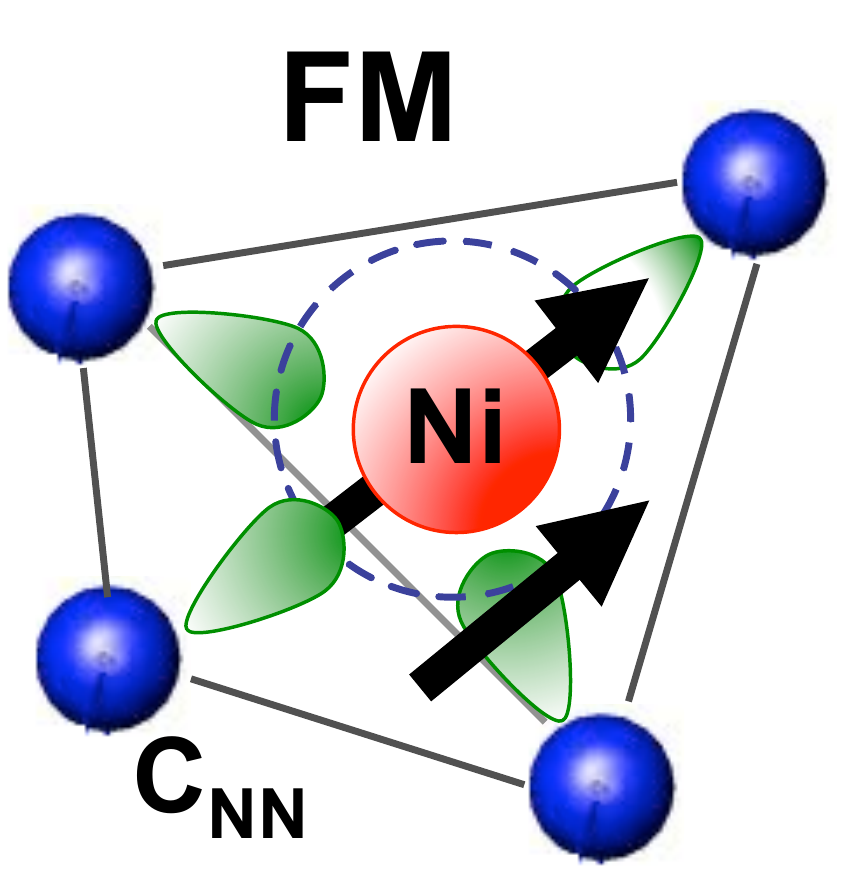} & \includegraphics[width =32mm]{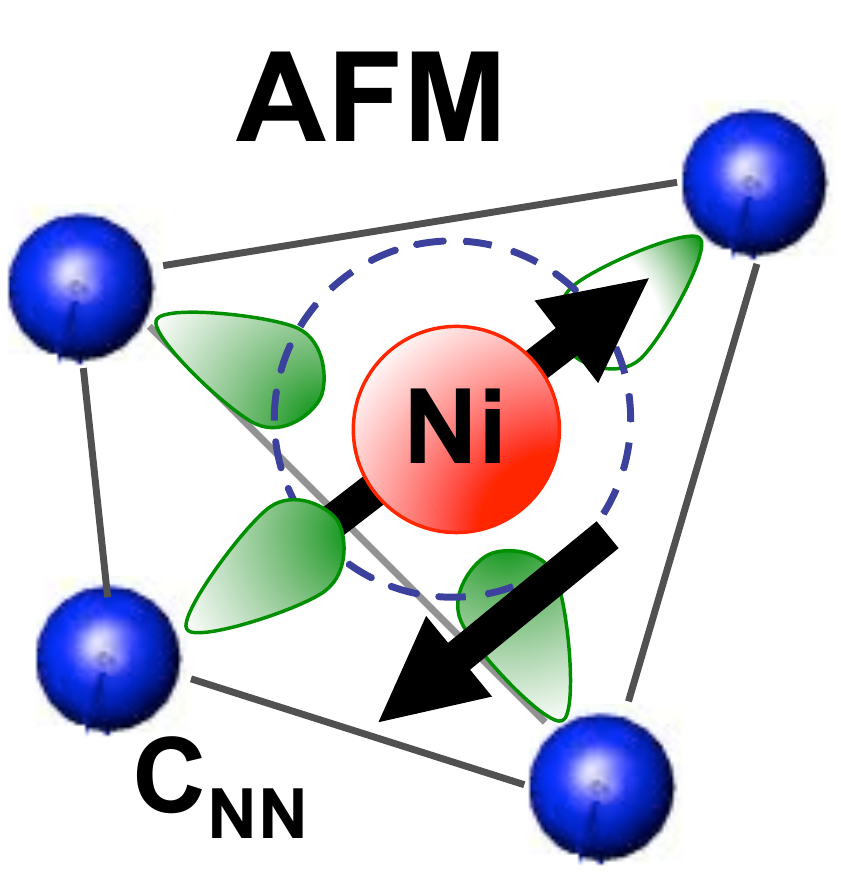}\\ & \\[-0.2cm] \hline 
      & \\[-0.2cm]
       \includegraphics[width = 38mm]{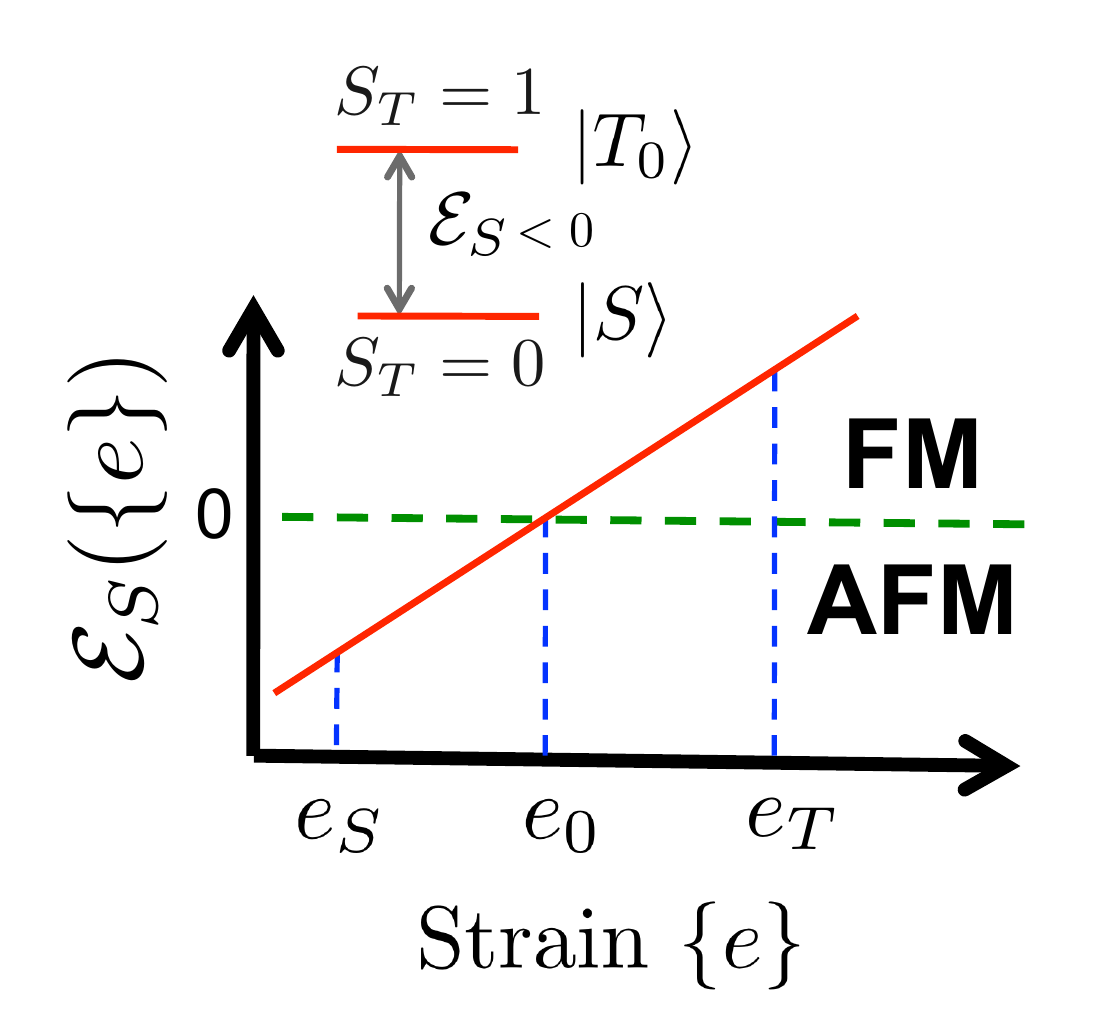} & \includegraphics[width = 38mm]{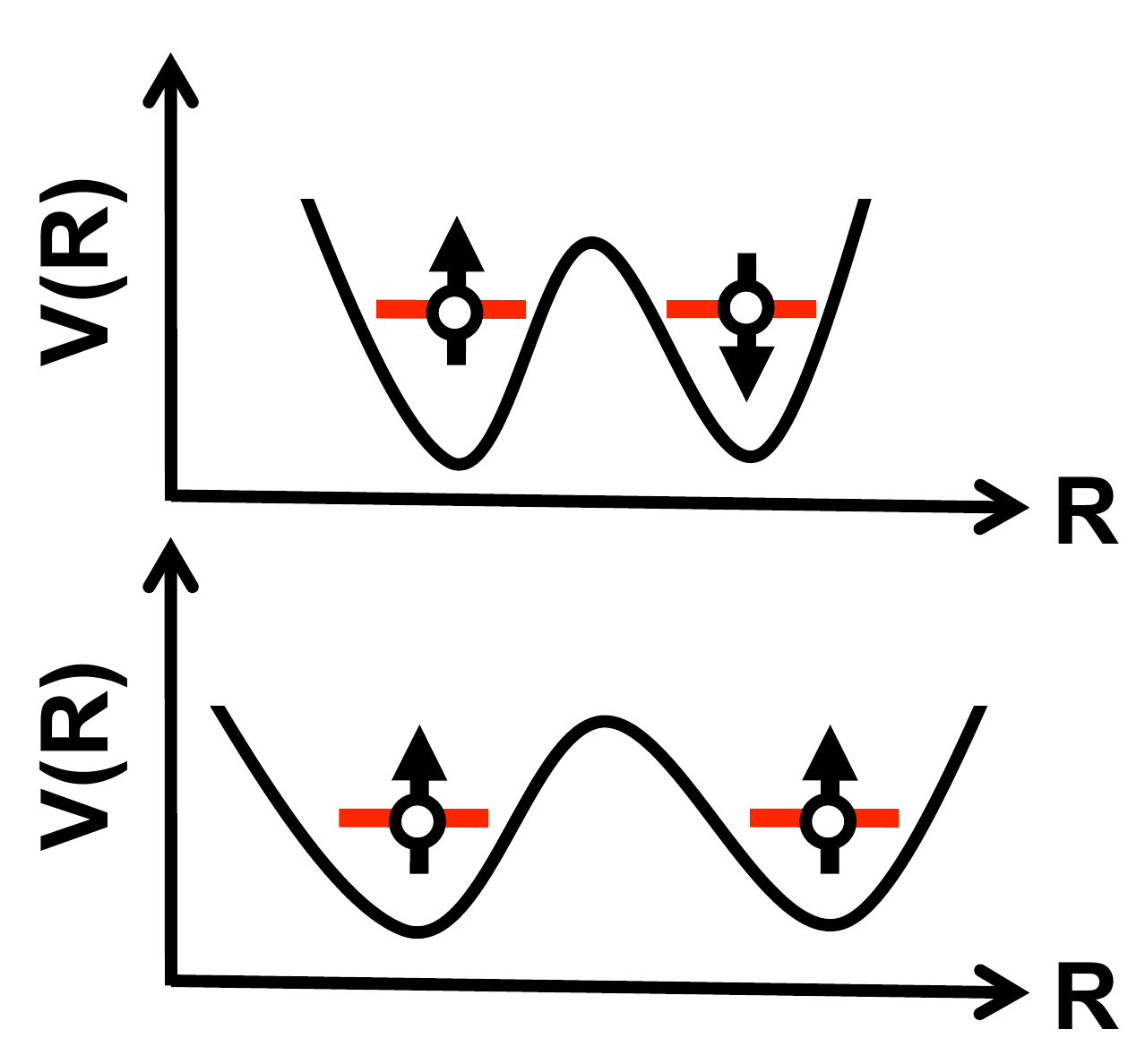} \\   \hline

\end{tabular}
\caption{Ni$^0_s$ qubit (top) : singlet $S$ (AFM) - triplet $T$ (FM) groundstate of the spins $S_1$ and $S_2$ (black arrows) depending on  strain.  
qubit initialization and potential $\mathrm{V(R)}$ for the two electronic spins as a function of their separation R (bottom).}\label{QD} 
\end{figure}

Fig. \ref{QD} shows (top) schematically the location and orientation of the spins in the Ni$^0_s$ qubit system. The Ni of spin $S_1=\frac{1}{2}$ is at the center of the tetrahedra formed by the four NN carbon atoms with a spin $S_2=\pm\frac{1}{2}$ distributed over the four NN carbon dangling bonds, describing a pseudo-orbit around the Ni atom.  Whereas in the simplest picture of quantum-dot gate control it is the barrier between the spins which is lowered, here the potential minima are brought closer together and the exchange splitting $\mathcal{E}_S$ is controlled by the application of strain (Eq. \ref{2stateham}). 

The qubit can be initialized  in the  singlet state ($e=e_{S}$, $S_T=0$) by the application of strain (Fig. \ref{QD}), or in the $T_+$  triplet state ($e=e_{T}$, $S_T=1$) in a static magnetic field. Figure~\ref{DOSnoStr} also reveals the potential of spin-selective optical processes to initialize the spin. Spin-orbit interaction within the partially-occupied spin-polarized $t_{2g}$ manifold of the Ni ion, not shown in the diagrams so far, will split the three states according to the orbital angular momentum projection of the $t_{2g}$ state ($+1$, $0$, $-1$) parallel to the spin, with the state with spin and orbit parallel at the highest energy (and thus unoccupied). Optical pumping with circularly-polarized light at about $3$~eV, corresponding to a transition from the valence maximum to the lowest-energy unoccupied minority-spin $t_{2g}$ feature in the density of states, will create minority spins in the $t_{2g}$ states if minority spin direction and light polarization are antiparallel, but not if parallel. Eventually, non-spin-selective recombination combined with spin-selective optical generation of carriers will drive the ground state spin $S_T=1$ to point antiparallel to the polarization of the optical pump. From the $T_+$ state a microwave pulse can coherently manipulate the spin into the $T_0$ state.  Once initialized, the qubit can be manipulated within the $S/T_0$ DFS by strain modulation and resonant microwave radiation\cite{Childress2006} through Eq.~(\ref{2stateham}). 

Readout of the quantum state of the Ni$^0_s$ qubit could be performed by electron paramagnetic resonance, or through spin-selective optical measurements. 
Because of the spin-orbit interaction described above, an optical probe with photon energy just below the valence to unoccupied $t_{2g}$ state transition will experience Faraday rotation that depends on the orientation of the Ni spin, which allows a measurement of the state of the system in the $|\uparrow_{\rm Ni}\downarrow_{\rm C}\rangle$, $|\downarrow_{\rm Ni}\uparrow_{\rm C}\rangle$ basis.  
 
In conclusion, we have performed {\it ab initio} calculations of the ground state of Ni$_s^0$ in diamond. The individual spins associated with this dopant, that of the Ni and of the nearest-neighbor C's, should each be limited in coherence time by similar processes in diamond that limit the coherence time of the NV$^-$ spin center, especially the presence of nuclear spins ($^{61}$Ni nuclear spins have similar rarity to $^{13}$C nuclear spins). 
Thus, due to the use of a decoherence-free subspace, a qubit architecture based on   Ni$^0_s$  may  be more immune from decoherence than a double-quantum-dot-based spin qubit or an NV$^{-}$ spin center in diamond. 

This work was supported by DARPA QuEST. We thank E. L. Hu and D. D. Awschalom for helpful discussions.

\bibliography{central-bibliography-3}

\end{document}